\documentclass{agujournal2019}
\usepackage{url} 
\usepackage{lineno}
\usepackage[finalnew]{trackchanges} 
\usepackage{soul}
\usepackage{nicefrac, physics, graphicx, txfonts}

\newcommand{\aap}{    {\it Astron. Astrophys. }}

\newcommand{\apjl}{   {\it Astrophys. J. Lett. }}
\newcommand{\apjs}{   {\it Astrophys. J. Suppl. S.}}

\journalname{JGR: Space Physics}

\begin{document}

%
%


\title{Writhed analytical magnetic flux rope model}

%
%




\authors{A. J. Weiss \affil{1,2,3}, T. Nieves-Chinchilla\affil{4,5}, C. M\"ostl\affil{1}, M. A. Reiss\affil{6}, T. Amerstorfer\affil{1,2}, R. L. Bailey\affil{7}}


\affiliation{1}{Austrian Space Weather Office, Zentralanstalt f\"ur Meteorologie und Geodynamik, Graz, Austria}
\affiliation{2}{Space Research Institute, Austrian Academy of Sciences, Schmiedlstraße 6, 8042 Graz, Austria}
\affiliation{3}{Institute of Physics, University of Graz, Universit\"atsplatz 5, 8010 Graz, Austria}
\affiliation{4}{Heliospheric Physics Laboratory, NASA Goddard Space Flight Center, Greenbelt, MD 20771, USA}
\affiliation{5}{Department of Physics, Catholic University of America, Washington, DC, USA}
\affiliation{6}{Community Coordinated Modeling Center, NASA Goddard Space Flight Center, Greenbelt, MD 20771, USA}
\affiliation{7}{Conrad Observatory, Zentralanstalt f\"ur Meteorologie und Geodynamik, Hohe Warte 38, 1190 Vienna, Austria}




\correspondingauthor{Andreas J. Weiss}{ajefweiss@gmail.com}




\begin{keypoints}
\item We develop an analytical model that can be used to describe writhed flux ropes by describing the flux rope axis as a general space curve.
\item We show how this model can be implemented numerically in terms of quadratic splines and configured for an arbitrary twist distribution.
\item We find that the field lines resulting from our model have lower twist per unit length than would be expected from a toroidal approximation.
\end{keypoints}

%
%

%
%


\begin{abstract}
Observations of magnetic clouds, within interplanetary coronal mass ejections (ICMEs), are \add{often} well described by \remove{magnetic} flux rope models. Most of these \remove{models} assume either a cylindrical or toroidal geometry. In some cases, \add{these} models are also capable of accounting for non-axisymmetric cross-sections but they generally all assume axial invariance. It can be expected that any ICME, and its flux rope, will be deformed along its axis due to \remove{external} influences such as the solar wind. \add{In this work}, we aim to develop a writhed analytical magnetic flux rope model which would allow us to analytically describe a flux rope structure with varying curvature and torsion so that we are no longer constrained to a cylindrical or toroidal geometry. In this first iteration of our model we will solely focus on a circular cross-section of constant size. We describe our flux rope geometry in terms of a parametrized flux rope axis and a parallel transport frame. We derive expressions for the axial and poloidal magnetic field components under the assumption that the total axial magnetic flux is conserved. \change{We present a fully analytical solution for the case of an arbitrarily curved flux rope within a plane. In the case of a writhed flux rope, with arbitrary curvature and torsion, we find an approximate solution that is valid for small curvature. In both cases we find that the twist of the magnetic field changes locally when the geometry deviates from a cylinder or torus.}{We find an entire class of possible solutions, which differ by the choice of integration constants, and present the results for a specific example. In general, we find that the twist of the magnetic field locally changes when the geometry deviates from a cylinder or torus. This new approach also allows us to generate completely new types of in situ magnetic field profiles which strongly deviate from those generated by cylindrical or toroidal models.}
\end{abstract}

\section{Introduction}
A magnetic flux rope is a confined magnetic field structure consisting of a flux tube and an axially twisted internal magnetic field. These structures play a prominent role in heliophysics and in many other astrophysical settings, and are believed to exist at the core of any interplanetary coronal mass ejection (ICME). The in-situ magnetic field measurements of these flux ropes structures within ICMEs were initially named magnetic clouds \cite{Burlaga_1981} before they were \remove{associated with coronal mass ejections (Gosling et al. 1991) and successively} found to closely follow the signature of a magnetic flux rope \cite{Goldstein_1983,Marubashi_1986,Bothmer_1998}.

The basic magnetic field structure of a flux rope can be described using cylindrical analytical models such as uniform-twist force-free models \cite{Gold_1960} or linear force-free configurations \cite{Lundquist_1950,Lepping_1990, Farrugia_1995}. \add{Other interpretations do exist, such as multi tube flux ropes} \cite<e.g.,>[]{Osherovich_1999}\add{, or spheromaks  } \cite<e.g.,>[]{Vandas_1997} \add{which we will not cover.} From in-situ magnetic field measurements and white light observations using coronagraphs and heliospheric imagers \cite<e.g.,>[]{Mulligan_2001, Vandas_2005, vourlidas_2013, Davies_2021} we know that these cylindrical approximations are highly simplified, and that the geometry of ICMEs can be significantly more complicated due to interaction with the coronal magnetic field \cite{Lugaz_2012, Kay_2015, moestl_2015} or the solar wind \cite{Riley_2004A, Liu_2006, Demoulin_2009}. These general deformations can be very hard to identify in the local in-situ magnetic field measurements and the measurements are also affected by other processes, such as flux rope expansion \cite{Leitner_2007, Gulisano_2012}. These problems are additionally exacerbated when only single spacecraft measurements are available.

Recent efforts have focused on constructing models with higher complexity regarding the geometry or the internal magnetic field structure with the aim of better reconstructing the measured in-situ signatures. These studies include purely analytical approaches \cite{Hidalgo_2002,Vandas_2017,Vandas_2017b,NC_2018}, and also semi-analytical models \cite{Isavnin_2016,kay_2018,Weiss_2021a}. One of the key components of any recently developed analytical model is axial invariance so that the basic geometry always corresponds to either a cylinder or torus and only the cross-sections are changed. This excludes the possibility of modelling any axial deformations which are expected to appear due to interaction with the ambient solar wind \cite<e.g.,>[]{Rollett_2014, Hinterreiter_2021}.

In this paper, we make use of the mathematical framework developed in \citeA{NC_2016,NC_2018} (henceforth referred to as NC16/NC18) and introduce a writhed, or bent, flux rope model that allows for arbitrary curvature and torsion. In Section \ref{sec:model}, we describe how we combine the approach in NC16/NC18 with the concept of a parallel transport frame, which allows us to build a continuous curvilinear coordinate system on top of a parametrized flux rope axis. We then derive expressions for the magnetic field components assuming that the axial flux is invariant along the flux rope. We also derive equations that allow for our model to be configured in terms of a predetermined twist distribution function in Section \ref{sec:configuration} when the flux rope is locally a cylinder. \change{Two}{An} exemplary \change{geometries}{writhed flux rope}  \remove{and related magnetic field solutions are }\add{is} then presented in Section \ref{sec:example} for illustration purposes.\add{ Here we show how the field lines behave differently compared to a more classical toroidal geometry and also show the magnetic field intensity, the twist and the Lorentz force by proxy at three different cross-section cuts}. A discussion of our approach, the results and an outline of how the model can be extended to more complicated geometries is then performed in Section \ref{sec:dc}.                                           
\section{Model} \label{sec:model}

Our goal is to build a flux rope model, with a circular cross-section, that can be arbitrarily bent along the axis. As such we cannot, as is commonly done in cylindrical models, align the flux rope axis with the $z$ \change{coordinate}{axis of the coordinate system}. Instead we describe the flux rope axis using an arbitrarily parametrized path $\vb*{\gamma}(s)$ \change{(where we will henceforth generally omit any dependencies on the $s$ coordinate to keep the notation cleaner)}{. We will henceforth generally omit any dependencies on the $s$ coordinate to keep the notation cleaner}. In the case of a classical cylindrical geometry the path $\vb*{\gamma}$ would be a straight line and in the case of a torus it would be a closed circle. We can then create a curvilinear coordinate system that describes our flux rope geometry:
\begin{linenomath*}
\begin{equation}
\label{eq:coordinate_system}
\vec{r}(r,s,\varphi)=\vb*{\gamma}-r\sigma\,\vb*{n}_1\cos\varphi-r\sigma\,\vb*{n}_2\sin\varphi
\end{equation}
\end{linenomath*}
where $r$ is the radial distance coordinate from the flux rope axis, $s$ the coordinate along the axis, $\varphi$ the azimuthal angle with respect to $\vb*{n}_{1/2}$ and $\sigma$ the fixed half-width of the flux rope. The two normal vectors $\vb*{n}_{1/2}$ are chosen so that 
$\{\vb*{t},\vb*{n}_1,\vb*{n}_2\}$ forms an orthonormal set where $\vb*{t}=\nicefrac{\partial_s\vb*{\gamma}}{\norm{\partial_s\vb*{\gamma}}}$ is the normalized velocity of $\vb*{\gamma}$. The volume of the flux rope is then defined by the coordinate range $r \in [0, 1]$, $s \in [0, s_\textrm{max}]$ and $\varphi \in [0, 2\pi)$. The vectors $\{\vb*{t}_1,\,\vb*{n}_1,\vb*{n}_2\}$ are also all functions of the coordinate $s$ without it explicitly being written. We further introduce a shorthand for the velocity of the path as $v=\norm{\partial_s\vb*{\gamma}}$.

We \change{will use}{make use of} the same mathematical framework as in NC18 but will not repeat the \remove{required} basic definitions as they are explained in detail \remove{and used} in NC18. We \remove{make} use of standard Einstein notation with upper and lower indices indicating contravariant and covariant quantities respectively and \change{raising or lowering of}{raise or lower} indices via contraction with the metric tensor. Any quantities described in our coordinate system with the non-unit basis vectors are denoted with a $c$ subscript and related to the scaled physical quantities via the appropriate scale factors. As in NC18 we start \change{with}{by} constructing the covariant basis vectors of our coordinate system. These are defined as $\vb*{\epsilon}_i = \partial_i \vec{r}(r,s,\varphi)$ and we are able to directly \change{compute}{evaluate expressions for} \remove{both} $\vb*{\epsilon}_r$ and $\vb*{\epsilon}_\varphi$ without issues. In order to find an expression for $\vb*{\epsilon}_s$ we require a more explicit description \change{for}{of} the vectors $\vb*{n}_{1/2}$ for which there are multiple approaches. The simplest approach makes use of the so-called Frenet-Serret vectors so that the normal vectors are defined as:
\begin{linenomath*}\begin{eqnarray}
\vb*{n}_1 = \vb*{n} &=& \flatfrac{\partial_s\vb*{t}}{\norm{\partial_s\vb*{t}}},\label{eq:fs_n}\\
\vb*{n}_2= \vb*{b} &=& \vb*{t}\cross\vb*{n}\label{eq:fs_b}.
\end{eqnarray}\end{linenomath*}
The Frenet-Serret vectors are also accompanied by the Frenet-Serret equations which describe the derivatives of $\{\vb*{t},\,\vb*{n},\,\vb*{b}\}$ with respect to the $s$ coordinate:
\begin{linenomath*}\begin{eqnarray}
\partial_s \vb*{t} &=& v\kappa \vb*{n},\label{eq:fs_eq_1}\\
\partial_s \vb*{n} &=& -v\kappa \vb*{t} + 
v\tau \vb*{b},\label{eq:fs_eq_2}\\
\partial_s \vb*{b} &=& -v\tau \vb*{n},\label{eq:fs_eq_3}
\end{eqnarray}\end{linenomath*}
where $\kappa$ is the curvature and $\tau$ is the torsion.
The \remove{most significant} issue with this approach is that the Frenet-Serret vectors are ill-defined for points on the path $\vb*{\gamma}$ where the curvature vanishes. At these positions along the curve, it is not possible to construct a Frenet-Serret frame. Another drawback is that the resulting basis vectors are not necessarily orthogonal so that an additional orthogonalization trick must be employed to simplify the resulting expressions and equations \cite<e.g.,>[]{Yeh_1986, Prior_2016}. \add{Under certain constraints for $\vb*{\gamma}$, one can make use of the concepts that are described in }\citeA{Carroll_2013}\add{ to rectify the core issues of the Frenet-Serret approach by introducing a signed curvature.}

\begin{figure}[ht!]
    \centering
    \includegraphics[trim=0 100 0 50, clip,width=.6\linewidth]{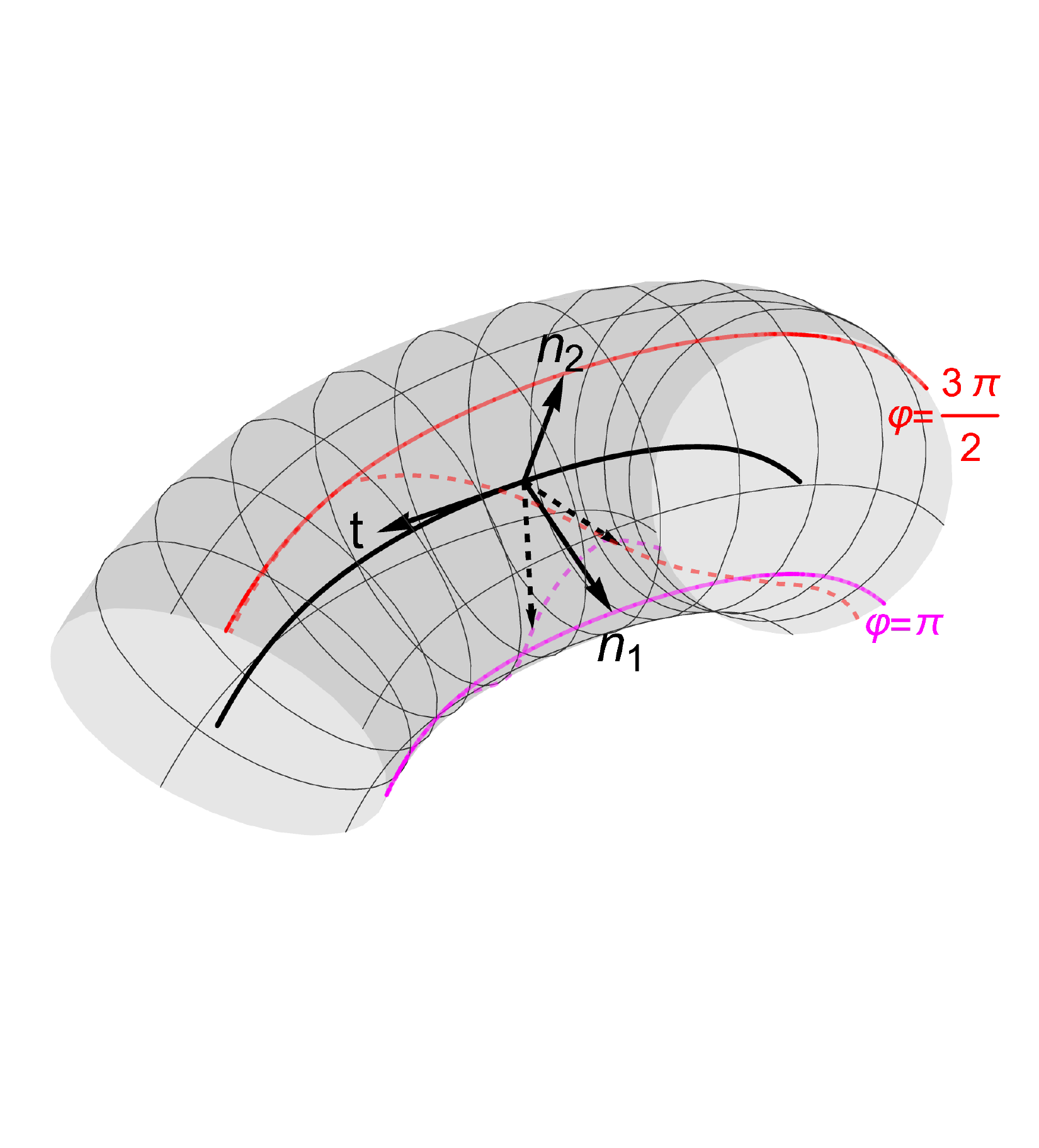}
    \caption{\label{fig:ptf}An arbitrarily curved flux rope example with a parallel transport frame. The solid arrows $\{\vb*{t},\,\vb*{n}_1,\vb*{n}_2\}$ show the basis vectors of the parallel transport frame at a specific section along a curve. The corresponding Frenet-Serret normal vectors $\{\vb*{n},\vb*{b}\}$, at the same location, are shown as dashed arrows. The two colored lines show $\vb*{\gamma}+\sigma\vb*{n}_{1/2}$ in magenta and red respectively and for both the parallel transport frame (solid) and the Frenet-Serret case (dashed). This demonstrates \change{that}{how} the two coordinate systems evolve differently along the axis.}
\end{figure}

\remove{For the above mentioned reasons we will instead use an alternative approach.} We \add{instead} make use of the concept introduced in \citeA{Bishop_1975}, that defines a parallel transport frame, \change{that}{which} is implicitly given by the equations:
\begin{linenomath*}
\begin{eqnarray}
\partial_s \vb*{t} &=&v k_1 \vb*{n}_1 +v k_2 \vb*{n}_2,\label{eq:ptf_eq_1}\\
\partial_s \vb*{n}_1 &=& -v k_1 \vb*{t},\label{eq:ptf_eq_2}\\
\partial_s \vb*{n}_2 &=& -v k_2 \vb*{t},
\label{eq:ptf_eq_3}
\end{eqnarray}
\end{linenomath*}
where $k_{1/2}$ are now a different set of curvature values. The resulting frame is sometimes also referred to as the Bishop frame or a relatively parallel adapted field. \add{Figure} \ref{fig:ptf} \add{shows an example of our flux rope geometry in terms of a parallel transport frame and also compares the normal vectors generated by parallel transport frame with the normal vectors created by the Frenet-Serret approach.} The values of $k_{1/2}$ are directly related to the Frenet-Serret curvature $\kappa$ via the relation $k_1^2+k_2^2=\kappa^2$. As is explained in \citeA{Bishop_1975}, there is no unique solution to these equations and there are in fact an infinite number of parallel transport frames with different solutions for $k_{1/2}$. All these parallel transport frames only differ \change{up to}{by} an angle of rotation. If we are given a solution $(\vb*{n}_1,\vb*{n}_2,k_1,k_2)$, \change{then $(\mathcal{R}_{\vb*{t}}(\psi)\vb*{n}_1,\mathcal{R}_{\vb*{t}}(\psi)\vb*{n}_2,\cos\psi k_1-\sin\psi\, k_2,-\sin\psi\, k_1+\cos\psi\, k_2)$ is also a valid solution where $\mathcal{R}_{\vb*{t}}(\psi)$ rotates a vector around $\vb*{t}$ by the angle $\psi$.}{we can write the totality of solutions as:}
\begin{linenomath*}\begin{equation}
(\cos\psi\,\vb*{n}_1-\sin\psi\, \vb*{n}_2,-\sin\psi\, \vb*{n}_1+\cos\psi\, \vb*{n}_2,\cos\psi\, k_1-\sin\psi\, k_2,-\sin\psi\, k_1+\cos\psi\, k_2)\label{eq:rpaf_totality}
\end{equation}\end{linenomath*}
\add{for any angle $\psi$.} There is sadly no easy way to construct the normal vectors, as with the Frenet-Serret approach, and the only way to find $\vb*{n}_{1/2}$ is to integrate Eqs. (\ref{eq:ptf_eq_1}-\ref{eq:ptf_eq_3}) from a specific starting point. \add{The relevant equations for this, that we used, are Eq.} \eqref{eq:ptf_eq_2} and:
\begin{linenomath*}
\begin{equation}
\partial_s k_{1/2} = v^{-1}\left(\vb*{n}_{1/2}\cdot\partial_s^2\vb*{t}-k_{1/2}\partial_s v\right),\label{eq:dkds}
\end{equation}
\end{linenomath*}
\add{where we then get the full solution by computing $\vb*{n}_2=\vb*{t}\cross\vb*{n}_1$.} This may appear to be problematic for implementing the model but we will show that this is not an issue for evaluating the magnetic field components. \remove{Figure shows an example of our flux rope geometry in terms of a parallel transport frame and also compares the normal vectors generated by parallel transport frame with the normal vectors created by the Frenet-Serret approach.}

\change{The metric tensor entries for our coordinate system $g_{ij}=\vb*{\epsilon}_i\cdot\vb*{\epsilon}_j$, can then be evaluated as:}{After constructing the covariant basis vectors, as in NC18, we can evaluate the metric tensor entries for our coordinate system $g_{ij}=\vb*{\epsilon}_i\cdot\vb*{\epsilon}_j$ as:}
\begin{linenomath*}
\begin{eqnarray}
g_{rr}=h_r^2&=&\sigma^2,\\
g_{rs}=g_{sr}&=&0,\\
g_{r\varphi}=g_{\varphi r}&=&0,\\
g_{ss}=h_s^2&=&v^2\left[1+r\sigma\cos\varphi\, k_1+r\sigma\sin\varphi\, k_2\right]^2,\\
g_{s\varphi}=g_{\varphi s}&=&0,\\
g_{\varphi\varphi}=h_\varphi^2&=&r^2\sigma^2,
\end{eqnarray}
\end{linenomath*}
where $h_i$ are the scale factors. We see that the basis vectors are orthogonal as only the diagonal entries are non-zero. The metric determinant $g$ takes the form:
\begin{linenomath*}
\begin{equation}
g= r^2 v^2 \sigma^4 \left[1+r\sigma\cos\varphi\, k_1+r\sigma\sin\varphi\, k_2\right]^2.
\end{equation}
\end{linenomath*}
This metric determinant must always be positive, which sets constraints for the $\sigma$ parameter and the curvature values $k_{1/2}$. We can show that $g>0$ for any $\varphi$ if, and only if:
\begin{equation}
\sigma^2\kappa^2=\sigma^2\left(k_1^2+k_2^2\right) < 1.
\end{equation}

\subsection{Magnetic Field Components}

We can then state the relevant equations that couple the contravariant magnetic field components $B_c^i$ and the contravariant current density components $j_c^i$ which are almost the same as in NC18 and are given by:
\begin{linenomath*}\begin{eqnarray}
0&=&\partial_s \left(g^{\nicefrac{1}{2}} B_c^s\right) + \partial_\varphi \left(g^{\nicefrac{1}{2}} B_c^\varphi\right),\label{eq:maxwell_div}\\
\mu_0 j^r_c&=&g^{-\nicefrac{1}{2}}\left[\partial_s \left(g_{\varphi\varphi} B_c^\varphi\right) - \partial_\varphi \left(g_{ss} B_c^s\right)\right],  \label{eq:maxwell_x}\\
\mu_0 j_c^s&=&-g^{-\nicefrac{1}{2}} \partial_r\left(g_{\varphi\varphi}B_c^\varphi\right)  ,\label{eq:maxwell_y}\\
\mu_0j_c^\varphi&=&g^{-\nicefrac{1}{2}}\partial_r \left(g_{ss} B_c^s\right)   ,\label{eq:maxwell_z}
\end{eqnarray}\end{linenomath*}
where we assume that the radial magnetic field component $B_c^r$ vanishes. The difference with respect to the Eqs. (11-14) in NC18 is that we must now account for two additional $\partial_s$ terms. We will now find solutions to this set of equations under the condition of axial magnetic flux conservation.

Without loss of generality we assume that there exists a point on our path $s=s_0$ where $\vb*{\gamma}$ is a straight line so that $\eval{k_{1/2}}_{s=s_0}=0$ and the respective derivatives also vanish. The geometry thus locally corresponds exactly to a cylinder as is described in NC16. The components of the magnetic field, using the Equation (20) from NC18, can thus be written as:
\begin{linenomath*}\begin{eqnarray}
B_c^s\big|_{s=s_0} &=& \frac{1}{v}\left(B_c^s\big|_{r=0} + \mu_0 \sigma^2\int_0^r \dd{r'} r' j_c^\varphi\big|_{s=s_0}\right),\label{eq:b_s_sol_nc18}\\
B_c^\varphi\big|_{s=s_0} &=& -\frac{\mu_0 v}{h^2 r^2} \int_0^r \dd{r'} r' j_c^s\big|_{s=s_0},
\label{eq:b_p_sol_nc18}
\end{eqnarray}\end{linenomath*}
where we additionally accounted for the $v$ factor. In NC18 the equations for the magnetic field were resolved by describing the current in terms of a radial power series. For our model we will alternatively use a decomposition based on shifted Legendre polynomials, as it can be shown that they have certain beneficial properties for our purposes. We write the axial and poloidal current as:
\begin{linenomath*}\begin{eqnarray}
j^s\big|_{s=s_0} &=& v j_c^s\big|_{s=s_0} =\sum_{m=0}^\infty \beta_m\,\frac{1}{r}\partial_r\left( r^2 \tilde{P}_m(r)\right) ,\\
j^\varphi\big|_{s=s_0} &=& r\sigma j_c^\varphi\big|_{s=s_0} = -\sum_{n=1}^\infty \alpha_n\,\partial_r\tilde{P}_n(r),
\end{eqnarray}\end{linenomath*}
where $\tilde{P}_i(r) = P_i(2r-1)$ are the shifted Legendre polynomials of $i$-th order that are defined for $r \in [0, 1]$. In contrast to NC18 the minimum value for the $n$-index now stems from the fact that $\partial_r P_0(r)=0$.
Evaluating the integrals in Eqs. (\ref{eq:b_s_sol_nc18}-\ref{eq:b_p_sol_nc18}) the magnetic field components then take the form:
\begin{linenomath*}\begin{eqnarray}
B_c^s\big|_{s=s_0} &=& \frac{1}{v}\left[B_c^s\big|_{r=0} - \mu_0 \sigma\sum_{n=1}^\infty\alpha_n \left(\tilde{P}_n(r) - \tilde{P}_n(0)\right)\right]\nonumber\\
&=&\underbrace{\frac{1}{v}B_c^s\big|_{r=0} + \frac{\mu_0\sigma}{v}\sum_{n=1}^\infty\alpha_n\tilde{P}_n(0)}_{=-\flatfrac{\mu_0\sigma\alpha_0\tilde{P}_0(r)}{v}}-\frac{\mu_0\sigma}{v}\sum_{n=1}^\infty\alpha_n \tilde{P}_n(r)\nonumber\\
&=&- \frac{\mu_0\sigma}{v}\sum_{n=0}^\infty\alpha_n \tilde{P}_n(r),\\
B_c^\varphi\big|_{s=s_0} &=& -\mu_0 \sum_{m=0}^\infty \beta_m \tilde{P}_m(r),
\end{eqnarray}\end{linenomath*}
where we additionally introduce the $\alpha_0$ coefficient for $\tilde{P}_0(r)=1$ to further simplify the expression and replace the $B_c^s\big|_{r=0}$ parameter. 
We continue by making the following ansatz for the general form of the axial magnetic field component:
\begin{linenomath*}\begin{equation}
B_c^s = A_s(r,s,\varphi)B_c^s\big|_{s=s_0},
\label{eq:gp_ansatz}
\end{equation}\end{linenomath*}
where $A_s$ is an auxiliary function that  fully encapsulates the axial and angular dependency of the general expression. We can directly solve for $A_s$ in the toroidal case with constant curvature $k_{1/2}=\textit{const.}$ and no radial current $j_c^r=0$. For this scenario Eq. \eqref{eq:maxwell_x} reduces to:
\begin{linenomath*}\begin{equation}
0 = g^{-\nicefrac{1}{2}}\partial_\varphi\left(g_{ss}\big|_{\substack{k_{1/2}=\textit{const.}}} A_s\big|_{\substack{k_{1/2}=\textit{const.}}}\right)B^s_c
\end{equation}\end{linenomath*}
for which we find that:
\begin{linenomath*}\begin{equation}
A_s\big|_{\substack{k_{1/2}=\textit{const.}}}= C(r,s) \left[1+r\sigma\cos\varphi\, k_1+r\sigma\sin\varphi\, k_2\right]^{-2}.
\end{equation}\end{linenomath*}
The integration constant $C(r,s)$ for this particular solution can be found by demanding conservation of the axial flux $\Phi^s $ for any constant values of $k_{1/2}$ with the same arrangement of coefficients. Due to the lack of any radial magnetic field component the axial flux must not only be conserved over the entirety of the cross section but also for \remove{each radial element} $\partial_r \Phi^s$. \change{We first compute the flux in the cylindrical case $\eval{\partial_r\Phi^s}_{s=s_0}$ which must be equal to the flux for the toroidal case so that we can directly infer $C(r,s)$:}{We first compute $\eval{\partial_r\Phi^s}_{s=s_0}$ in the cylindrical case which must be equal to the same expression for the toroidal case which allows us to directly infer $C(r,s)$:}
\begin{linenomath*}\begin{eqnarray}
\eval{\partial_r\Phi^s}_{s=s_0} &=&  \int_0^{2 \pi}\dd{\varphi} g^{\nicefrac{1}{2}} B_c^s\big|_{s=s_0} = 2 \pi r \sigma^2 vB_c^s\big|_{s=s_0}\nonumber \\
&=&\int_0^{2 \pi}\dd{\varphi}  \frac{g^{\nicefrac{1}{2}}C(r,s)\, B_c^s\big|_{s=s_0}}{\left[1+r\sigma\cos\varphi\, k_1+r\sigma\sin\varphi\, k_2\right]^2}\nonumber\\
&=&   \frac{2\pi r \sigma^2\, C(r,s)\, v B_c^s\big|_{s=s_0}}{\sqrt{1 - r^2\sigma^2\left(k_1^2+k_2^2\right)}}
\end{eqnarray}\end{linenomath*}
\begin{linenomath*}\begin{eqnarray}
\quad\implies C(r,s) &=& \sqrt{1 - r^2\sigma^2\left(k_1^2+k_2^2\right)}\\
\quad\implies A_s\big|_{\substack{k_{1/2}=\textit{const.}}} &=& \frac{\sqrt{1 - r^2\sigma^2\left(k_1^2+k_2^2\right)}}{\left[1+r\sigma\cos\varphi\, k_1+r\sigma\sin\varphi\, k_2\right]^2}
\end{eqnarray}\end{linenomath*}
Given our solution for the axial field in the cylindrical or toroidal case we can now note that $g^{\nicefrac{1}{2}}$ only depends on the curvature values $k_{1/2}$. As such our previously derived expression for $A_s$ conserves the axial flux regardless of how the flux rope is curved or twisted. We can thus use the existing toroidal expression for the axial magnetic field for the general case and assume that the poloidal field and the current conform so that the Eqs. (\ref{eq:maxwell_div}-\ref{eq:maxwell_z}) are resolved. For the general case it is important to keep in mind that we could also alternatively use different solutions of the form $A_s+\mathcal{C}_s$ where $\int_0^{2\pi}\dd{\varphi}\mathcal{C}_s=0$ and $\eval{\mathcal{C}_s}_{s=s_0}=0$. \add{The constraint $\eval{\mathcal{C}_s}_{s=s_0}=0$ is technically not necessary as we could introduce $\mathcal{C}_s$ in the same way for classical cylindrical or toroidal models. As a consequence it would follow that $j^r \neq 0$ and the cylindrical or toroidal flux rope would lose azimuthal symmetry. In the standard approach it is assumed that $j^r = 0$, which implicitly sets $\eval{\mathcal{C}_s}_{s=s_0}=0$. But in our scenario we make no assumptions for the radial current so that any argument for a specific choice for $\mathcal{C}_s$ appear weak. We plan to investigate the consequences of specific choices for $\mathcal{C}_s$ in the future, and for now use the simplest approach and set $\mathcal{C}_s=0$ for this paper.}\remove{This modification only adds additional complexity so we will only use the simplest form with $\mathcal{C}_s=0$ for this paper.} For the axial magnetic field component, accounting for the scale factor, we can thus write:
\begin{linenomath*}\begin{eqnarray}
B^s &=&  -\frac{\mu_0\sigma\,\sqrt{1 - r^2\sigma^2\left(k_1^2+k_2^2\right)}}{1+r\sigma\cos\varphi\, k_1+r\sigma\sin\varphi\, k_2}\sum_{n=0}^\infty\alpha_n \tilde{P}_n(r),\label{eq:bs_result}
\end{eqnarray}\end{linenomath*}
Applying this assumption to Eq. \eqref{eq:maxwell_div} and further assuming that $B_c^\varphi = A_\varphi(r,s,\varphi)\eval{B_c^\varphi}_{s=s_0}$ we can \remove{directly} reconstruct the poloidal field component, which takes the form:
\begin{linenomath*}\begin{eqnarray}
B^\varphi &=& -\frac{\mu_0 r \sigma}{h_s} \sum_{m=0}^\infty \beta_m \tilde{P}_m(r) - Q_\varphi\frac{\mu_0 r \sigma}{h_s}\sum_{n=0}^\infty\alpha_n \tilde{P}_n(r),\label{eq:bp_result}\\
Q_\varphi&=& \frac{r\sigma^2\left(\left[\sin\varphi +r\sigma k_2\right]\partial_s k_1-\left[\cos\varphi + r\sigma k_1\right]\partial_s k_2\right)}{v\, \left(1+r\sigma\cos\varphi\, k_1+r\sigma\sin\varphi\, k_2\right) \sqrt{1 - r^2\sigma^2\left(k_1^2+k_2^2\right)}}-\mathcal{C}_\varphi
\end{eqnarray}\end{linenomath*}
\add{where $\mathcal{C}_\varphi$ is the integration constant that we set to zero. As with $\mathcal{C}_s$, other choices for $\mathcal{C}_\varphi$ are also possible which will be investigated in the future.}
\remove{We find two additional terms in the poloidal field that depend on the poloidal current coefficients $\alpha_n$. The expression in Eq. has a big issue, namely the last term, which diverges if $k_1 \rightarrow 0$. In the case that the flux rope is confined to a plane, we can select a specific parallel transport frame in which $k_2=\partial_s k_2 = 0$ everywhere along the path so that the $k_1^{-1}$ does not appear when solving the equations. The poloidal magnetic field component, accounting for scale factors, can then be written as:}
\remove{If $k_1=0,$ and $k_2\neq0$ we can always rotate the frame into another parallel transport frame so that $k_1\neq 0$. But this trick fails if the Frenet-Serret curvature $\kappa$ completely vanishes. This means that Eq. cannot be used for the general case, as the boundary conditions at and near $\eval{\vb*{\gamma}}_{s_0}$ are not valid for how we configure our model. Alternatively, we can build an approximate solution. For small $k_{1/2}$, we expand $B_c^s$ in terms of $k_{1/2}$ and then solve for the extra terms in $B_c^\varphi$ using the divergence equation. The first order approximate solution for $B^\varphi$, that is valid for any combination of small $k_{1/2}$, then takes the form:}
\remove{where we no longer have any issues with vanishing $k_1$.Note that any physical quantity should also be independent of the parametrization used for $\vb*{\gamma}$ and therefore independent of $v$. The $v$ factor shows up in these equations because the derivatives with respect to the $s$ coordinate also depend on the parametrization. Terms of the form $v^{-1}\partial_s k_{1/2}$ are not affected by the particular choice of parametrization.} \add{The result for $B^\varphi$ shows that, if and only if the derivatives of $k_{1/2}$ are non-zero, additional terms appear for the poloidal field which are dependent on the poloidal current coefficients.} The physical interpretation of this result is that the flux rope twist will locally change according to changes in the curvature. Depending on the sign of $k_{1/2}$, and their respective derivatives, the  twist can either increase or decrease and it is also further dependent on the $\varphi$ coordinate.

\subsection{Model Properties} \label{sec:physics}

By construction, the condition of current conservation is always fulfilled as long as the current is physical. This may not be the case due to singularities in the current which, due to our chosen description, can only appear at the center of the flux rope structure. An example is the poloidal current $j^\varphi\big|_{s=s_0}$ which must vanish at $r=0$. This condition can be shown to be equivalent to:
\begin{linenomath*}\begin{equation}
\sum_{n=1}^\infty \alpha_n \eval{\left(\partial_r \tilde{P}_n(r)\right)}_{r=0}=\sum_{n=1}^\infty (-1)^{n+1} (n^2 + n)\,\alpha_n=0.\label{eq:alpha_constraint}
\end{equation}\end{linenomath*}
which sets a constraint on the values for $\alpha_n$. No such constraint exists for the $\beta_m$ coefficients as $j^s$ can take non-zero values at $r=0$. By resolving Eqs. (\ref{eq:maxwell_x}-\ref{eq:maxwell_z}) we could also generate the expressions for the current density components in our flux rope model. Unfortunately these expressions do not have an easily tractable form and it is very hard to extract general statements on their structure. An important property is that generally $j_r^c \neq 0$ if the curvature changes. This means that the total amount of current within the flux rope will change over the axis due to changes in the curvature.

In the construction of our model we used the fact that the axial magnetic flux is constant throughout the structure. This axial flux can be evaluated as:
\begin{linenomath*}\begin{eqnarray}
\Phi^s&=&\int_0^1\int_0^{2\pi}\dd{r}\dd{\varphi}g^{\nicefrac{1}{2}}B_c^s=-2\pi\mu_0\sigma^3\sum_{n=0}^\infty\alpha_n\int_0^1\dd{r} r \tilde{P}_n(r)\nonumber\\
&=&-\pi\mu_0\sigma^3\left(\alpha_0+\frac{\alpha_1}{3}\right)
\label{eq:axflux}
\end{eqnarray}\end{linenomath*}
and we find that all terms with coefficients $n\geq2$ vanish. The axial flux is thus only dependent on the first two coefficients. Due to our definition of $\alpha_0$ there is a hidden interdependence with respect to all higher order coefficients, the relevance of which depends on how the model is configured.

The same calculation can also be performed for the poloidal flux, where we find that:
\begin{linenomath*}\begin{eqnarray}
\Phi^\varphi &=& \int_0^1 \int_0^{s_\textrm{max}}\dd{r}\dd{s}g^{\nicefrac{1}{2}}B_c^\varphi=-L\, \mu_0\sigma^2 \left(\frac{\beta_0}{2} + \frac{\beta_1}{6}\right)\nonumber\\
&&-\int_0^1 \int_0^{s_\textrm{max}}\dd{r}\dd{s}\left[\hdots\right]\sum_{n=0}^\infty\alpha_n\tilde{P}_n(r),
\label{eq:polflux}
\end{eqnarray}\end{linenomath*}
where $L$ is the total length of the path $\vb*{\gamma}$ and the two
latter terms from the poloidal magnetic field are written in an extremely compact form. It must be that the total flux is independent of the $\varphi$ coordinate and the latter integral terms in Eq. (\ref{eq:polflux}) should thus disappear. We can roughly show by expansion and integration by parts that these latter terms cancel if the flux rope is closed. The total poloidal flux is then only determined by the first two coefficients $\beta_0$ and $\beta_1$.

We also briefly take a look into the magnetic energy stored within the flux rope, which can be written as:
\begin{linenomath*}\begin{eqnarray}
E &=& \int_0^1\int_0^{s_\textrm{max}}\int_0^{2\pi}\frac{g_{ij}B_c^iB_c^j}{2\mu_0}\sqrt{g}\dd{r}\dd{s}\dd{\varphi}\nonumber\\&=&\int_0^1\int_0^{s_\textrm{max}}\int_0^{2\pi}\frac{g_{ss}\left(B_c^s\right)^2+g_{\varphi\varphi}\left(B_c^\varphi\right)^2}{2\mu_0}\sqrt{g}\dd{r}\dd{s}\dd{\varphi}\label{eq:energy_integral}
\end{eqnarray}\end{linenomath*}
Without explicitly computing this integral, we can show that a flux rope with a left-handed twist will have a different total energy than a right-handed twist. The twist of a flux rope in our model can be flipped by swapping the sign of all $\beta_m$ coefficients. Assuming a cylindrical or toroidal geometry the $\left(B_c^\varphi\right)^2$ quantity will be unchanged. The extra terms in Eq. \eqref{eq:bp_result} do not change under such a change of twist so that $\left(B_c^\varphi\right)^2$  will be different for a left-handed and right-handed flux rope and the terms will also not all cancel out during integration. As a result we can deduce that a left-handed and right-handed flux rope, assuming the same geometry will also behave or evolve differently. We will see this more concretely when examining the involved Lorentz forces. The three contravariant components of the Lorentz force can be calculated using the same Equations from Eq. (29) in NC18:
\begin{linenomath*}\begin{eqnarray}
F_c^r &=& g^{rr}\sqrt{g}\left(j_c^s B_c^\varphi-j_c^\varphi B_c^s\right)\\
F_c^s &=& \sqrt{g}\left(g^{s\varphi}j_c^r B_c^s-g^{ss}j_c^rB_c^\varphi\right)\\
F_c^\varphi &=& \sqrt{g}\left(g^{\varphi\varphi}j_c^rB_c^s-g^{s\varphi}j_c^r B_c^\varphi\right)
\end{eqnarray}\end{linenomath*}
where now in general all three values will be non-zero as there is a radial current. Nonetheless the typical arising radial currents will be comparatively small and the radial Lorentz force can be expected to be the dominant component. The expressions for these forces will be of similar complexity to those of the current and it is therefore not practical to show their full form. We can alternatively attempt to compute the net radial force that acts on a slice of our flux ropes by calculating:
\begin{linenomath*}\begin{equation}
<F^r> = \int_0^1\int_0^{2\pi}\dd{r}\dd{\varphi}\sqrt{g}\vb*{\epsilon}_r F_c^r.
\end{equation}\end{linenomath*}
Since we are only looking at a specific position along the axis, we can choose a specific parallel transport frame so that the normal vectors are aligned with the Frenet-Serret vectors. As such $k_1=\kappa$ and $k_2=0$, which simplifies some of the calculations. We compute $<F^r>$ by only taking $k_1$ into account up to third order, current coefficients up to first order, and drop any terms of $\partial_s k_{1/2}$ beyond first order, including mixed terms, so that we get:
\begin{linenomath*}\begin{eqnarray}
<F^r> &\approx& -\frac{\mu _0 \pi }{8}  \beta_0^2  \sigma ^4 k_1 \vb*{n}_1 v \left(5 \sigma ^2 k_1{}^2+6\right)
+\frac{ \mu _0 \pi}{4}   \alpha_0^2 \sigma ^6 k_1{}^3 \vb*{n}_1 v\nonumber\\
&&+\frac{ \mu _0 \pi}{12}  \alpha _0 \beta _0\sigma ^5\partial_s k_1\vb*{n}_2 \left(19 \sigma ^2 k_1{}^2+15 \right)\nonumber\\
&&-\frac{\mu_0\pi}{12}  \alpha _0 \beta _0 \sigma ^5 \partial_s k_2\vb*{n}_1 \left(13 \sigma ^2 k_1{}^2+15\right).
\label{eq:lorentz_force_net_radial}
\end{eqnarray}\end{linenomath*}
The second term $\alpha_0^2$ points inwards and thus acts as the tension force with the \change{first}{other} term given by $\beta_0^2$ pointing outwards and thus representing the magnetic hoop force. The two \change{latter}{last} terms appear if there is a change in curvature and also include $\vb*{n}_2$ so that the Lorentz force points out of the plane of curvature. The conclusion of this result is that the flux rope will undergo a writhing motion if there is any change in curvature \add{(specifically for $k_1$ when $k_2=0$)}. As this force contains terms of $\alpha_0\beta_0$ it is also dependent on the handedness of the magnetic field so that a left-handed flux rope will undergo a writhing motion in the opposite direction than a right-handed flux rope.

\subsection{Configuration of Coefficients} \label{sec:configuration}

We will now discuss how we can configure our model coefficients $\alpha_n$ and $\beta_m$ so that the resulting flux rope exhibits certain properties. As our coefficients are defined for a point $s_0$ where the flux rope is locally a cylinder it also makes sense to configure the model at the same position. We define an arbitrary twist function $Q(r)$ and demand that:
\begin{linenomath*}\begin{equation}
Q(r)=\sum_{l=0}^\infty \zeta_l \tilde{P}_l(r) = \frac{\eval{B^\varphi}_{s=s_0}}{r \eval{B^s}_{s=s_0}} = \frac{h_\varphi \eval{B_c^\varphi}_{s=s_0}}{r h_s \eval{B_c^s}_{s=s_0}} = \frac{\sum_{m=0}^\infty\beta_m \tilde{P}_m(r)}{\sum_{n=0}^\infty\alpha_n \tilde{P}_n(r)},
\label{eq:twist}
\end{equation}\end{linenomath*}
where $\zeta_l$ are the coefficients used for the expansion of $Q(r)$ in terms of shifted Legendre polynomials. Using the following two properties for shifted Legendre polynomials:
\begin{linenomath*}\begin{eqnarray}
\int_0^1\tilde{P}_m(r)\tilde{P}_n(r) &=&\frac{\delta_{nm}}{2m+1}\\
\int_0^1 \dd{r} \tilde{P}_{m+n-2k}(r)\tilde{P}_m(r)\tilde{P}_n(r) &=& \frac{1/2}{2m+2n-2k+1}\Lambda,\\ 
\Lambda &=& \frac{\lambda_k\lambda_{m-k}\lambda_{n-k}}{\lambda_{m+n-k}}\\\lambda_k &=& \frac{(2k)!}{2^k(k!)^2}
\end{eqnarray}
\end{linenomath*}
where a derivation for the integral over the triple product is given in \citeA{Dougall_1953}, we can rearrange Eq. \eqref{eq:twist} and solve for $\beta_m$ so that:
\begin{linenomath*}\begin{eqnarray}
\beta_m &=& (2m+1)\int_0^1 \dd{r} \tilde{P}_m(r)\left(\sum_{l=0}^\infty \zeta_l \tilde{P}_l(r)\right)\left(\sum_{n=0}^\infty \alpha_n \tilde{P}_n(r)\right)\nonumber\\
&=& (2m+1)\sum_{n=0}^\infty\sum_{k=0}^{\textrm{min}(m,n)} \frac{\alpha_n\zeta_{n+m-2k}}{2m + 2n - 2k + 1} \frac{\lambda_k\lambda_{m-k}\lambda_{n-k}}{\lambda_{m+n-k}}.\label{eq:twist_beta}
\end{eqnarray}\end{linenomath*}
Using Eq. \eqref{eq:twist_beta} we can thus configure our $\beta_m$ coefficients so that the resulting flux rope, in a cylindrical configuration, exhibits exactly the prescribed twist distribution $Q(r)$. The $\alpha_i$ coefficients must be determined using other constraints. This could be done using the Lorentz force, where at $s_0$ only $F_c^r$ is non-zero, and:
\begin{linenomath*}\begin{eqnarray}
\eval{F^r_c}_{s=s_0} &=& 
-\mu_0\sum_{n=0}^\infty\sum_{n'=0}^\infty\tilde{P}_n(r)\left(\partial_r\tilde{P}_{n'}(r)\right)\alpha_n\alpha_{n'}\nonumber\\
&&-2r\mu_0\sum_{m=0}^\infty\sum_{m'=0}^\infty\tilde{P}_m(r)\tilde{P}_{m'}(r)\beta_m\beta_{m'}\nonumber\\
&&-r^2\mu_0\sum_{m=0}^\infty\sum_{m'=0}^\infty\tilde{P}_m(r)\left(\partial_r\tilde{P}_{m'}(r)\right)\beta_m\beta_{m'}.
\end{eqnarray}\end{linenomath*}
We have not been able to find an iterative solution for evaluating the $\alpha_n$ coefficients according to a given force distribution $\eval{F^r_c}_{s=s_0}$. When using a limited number of coefficients the corresponding $\alpha_n$ coefficients can be easily found by using numerical minimization algorithms. For a force-free distribution we can minimize the integral $\int\dd{r} \left(\eval{F^r_c}_{s=s_0}\right)^2$, under the constraint of a given axial flux, where the integrand is a polynomial of order $2n+2m+2$.

We can compare our approach to another model for arbitrarily twisted flux ropes given in \citeA{Vandas_2019}. In their case an integral over the twist distribution function, and its derivatives, must be computed for evaluating the magnetic field at every point. In our case the calculations for the coefficients are done before hand and the evaluation of the model, given the coefficients, is then a simple polynomial. We can also fine tune the accuracy of our model by using more or less coefficients with more coefficients delivering a more force-free flux rope if this is desired.

Later on, when presenting a flux rope example we will make use of the simplest twist configuration possible and use a uniformly twisted field with only the $\zeta_0$ coefficient being non-zero. In this case it is easy to verify from Eq. \eqref{eq:twist_beta} that $\beta_i = \zeta_0\alpha_i$ for all indices. We also make use of the force-free condition which we can satisfy by calculating:
\begin{linenomath*}\begin{equation}
\alpha_i=(2i+1)\int_0^1\dd{r}\frac{B_c^s\big|_{r=0}}{1+\zeta_0^2r^2}\tilde{P}_i(r),
\label{eq:gold_hoyle_coefficient}
\end{equation}\end{linenomath*}
where we used the analytical form of the uniform-twist solution \cite{Gold_1960}. The number of coefficients that are required to deliver sufficient accuracy to approximate the Gold-Hoyle solution depends on the twist parameter $\zeta_0$ with a higher twist requiring higher order coefficients. Normally only few coefficients are needed for a good approximation of the magnetic field but significantly higher orders are required for accurately depicting the current (and therefore also the Lorentz forces). There also exists an optimal or maximum order beyond which the approximation will begin to diverge that is also further dependent on the twist. For our purposes we will use a dozen coefficients in order to deliver sufficient accuracy for the current and the derived Lorentz forces. It is at this point that we would also like to highlight why we have chosen our particular approach with the more complicated Legendre polynomials instead of using a radial power series expansion. A Taylor expansion of the Gold-Hoyle solution takes the following form:
\begin{linenomath*}\begin{equation}
\frac{1}{1+\zeta_0^2r^2} = 1 - \zeta_0^2r^2 + \zeta_0^4r^4 -\zeta_0^6r^6+\zeta_0^8r^8\hdots
\end{equation}\end{linenomath*}
where it is immediately apparent that the power series diverges for any twist values with $\zeta_0\geq1$. The models described in NC16/NC18 are thus incapable of describing all uniform twist configurations even with an unlimited number of coefficients. \remove{The same problem will likely persistent for any configuration with non-zero twist within the core.}

\subsection{Model Implementation}\label{sec:impl}

In contrast to cylindrical or toroidal flux rope models there is no general straightforward way to transform Cartesian coordinates into our curvilinear coordinates $(r,s,\varphi)$ and not all transformations are necessarily unique. By default uniqueness is guaranteed within the flux rope volume if the flux rope volume does not self-intersect anywhere. If the path $\vb*{\gamma}$ is given by a purely analytical function one can transform the coordinates using standard minimization algorithms. For numerical implementations, the path $\vb*{\gamma}$ can be implemented as a spline \cite<e.g.,>[]{Titov_2021} or a B\'ezier curve. We will focus on the spline implementation as it is the simplest. The problem of finding $(r,s)$ is equivalent to finding the closest point on a spline, which can be solved efficiently using numerical algorithms \cite<e.g.,>[]{Wang_2002}. Once $(r,s)$ is found the $\varphi$ coordinate depends on the specific orientation of the normal vectors. For the parallel transport frame we would normally have to define $\eval{\vb*{n}_{1/2}}_{s=s_0}$ and solve for $\vb*{n}_{1/2}$, and also the $k_{1/2}$ values, according to the Eqs. (\ref{eq:ptf_eq_1}-\ref{eq:ptf_eq_3}), which would allow us to determine $\varphi$. In practice, we can make use of the fact that we do not need to use the same frame everywhere along the curve and for any point on the curve we can use an arbitrary set of $\vb*{n}_{1/2}$ vectors as long as they satisfy the orthogonality condition. We can either use Frenet-Serret normal vectors, when possible, or generate one of the vectors using the formula $n_1 = \nicefrac{\vb{r}\cross\vb*{t}}{\norm{\vb{r}\cross\vb*{t}}}$ if $\vb{r}\cross\vb*{t} \neq 0$. When doing so the curvature values $k_{1/2}$ are not known, but can be calculated by applying the dot product in Eq. \eqref{eq:ptf_eq_1}\remove{and its respective derivative}:
\begin{linenomath*}\begin{equation}
k_{1/2} = v^{-1}\vb*{n}_{1/2}\cdot\partial_s\vb*{t}.
\end{equation}\end{linenomath*}
Together with Eq. \eqref{eq:dkds} this is all that is needed to evaluate the magnetic field. Note that for evaluating the current and the Lorentz forces we will also need $\partial_s^2 k_{1/2}$ which can also be found using the same approach. In the remaining parts of our paper we will always show the flux rope examples in terms of the same parallel transport frame, as stitching multiple frames together does not make sense for visualizations. But it is important that in terms of flux rope modeling, the evaluation of the magnetic field is much simpler as one does not need to use the same parallel transport frame everywhere.

\section{Exemplary Flux Rope} \label{sec:example}

\begin{figure*}[ht!]
\center
\includegraphics[trim=15 10 15 35, clip,width=.8\linewidth]{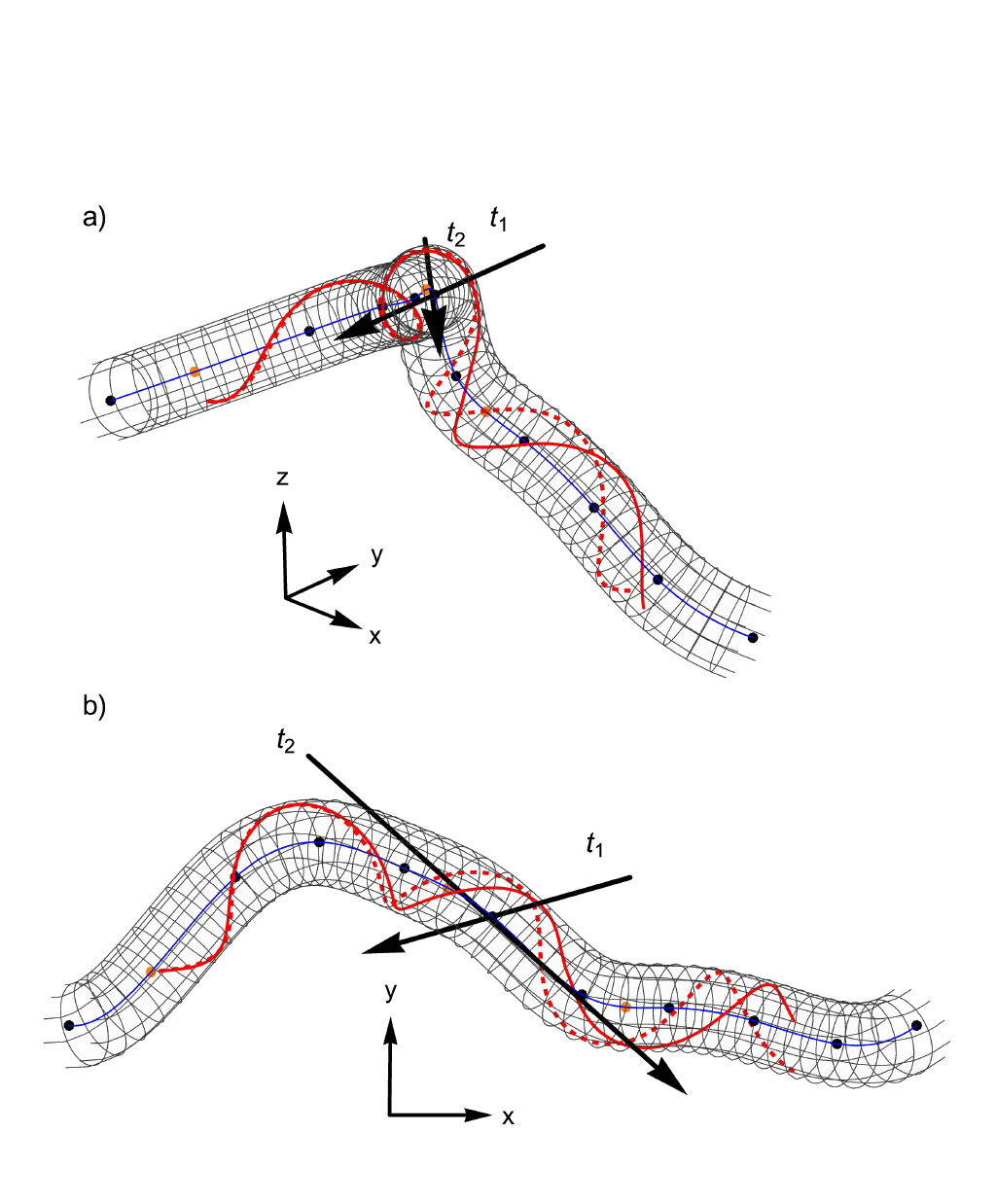}
\caption{\label{fig:exemplary_flux_rope} Exemplary flux rope structure generated using the spline approach with 11 control nodes (black) as seen from the side (a) and the top (b). The blue central line follows the parametrized path $\vb*{\gamma}$ and the two red lines are integrated magnetic field lines. The solid red field line gives the solution according to Eqs. (\ref{eq:bs_result}, \ref{eq:bp_result}), while the dashed red field line is based on a toroidal approximation which does not take account for effects of changing curvature values. Both field lines start at the same position $(r,s,\varphi) = (0.9, 0.1, \nicefrac{3\pi}{4})$ and the twist is set to $\zeta_0=-1$. The position $s=0$ corresponds to the left and $s=1$ to the right side in these figures. The two black lines and arrows, $t_1$ and $t_2$, represent two virtual spacecraft trajectories that we will later use to generate synthetic in situ profiles.}
\end{figure*}

We now explore our model using a \change{cubic}{quadratic} spline implementation. We set $\sigma=0.1$, $\eval{B_c^\varphi}_{r=0}=15\,\textrm{nT}$ and $\zeta_0=-1$ which serve as typical parameters for an ICME at $1~\textrm{au}$. Figure \ref{fig:exemplary_flux_rope} shows a side (a) and top (b) view  an exemplary flux rope structure with a specific parallel transport frame. The path $\vb*{\gamma}$ is described by a spline with 11 control nodes and 10 piece-wise \change{cubic}{quadratic} polynomial curves parametrized within the range $s\in [0,1]$. The control nodes are shown as black points and the interpolated spline $\vb*{\gamma}$ is shown as the blue line. The parametrization of the curve is left to right. The two red lines represent integrated magnetic field lines with the same starting position at $s=0.1$. The solid red line represents our solution from Eqs. (\ref{eq:bs_result}, \ref{eq:bp_result}). The dashed red line is a naive solution, where we set $\partial_s k_{1/2}=0$, so that this result can be seen as a toroidal approximation. As is shown in this example, the local twist for the field line changes significantly so that the total number of turns in our solution is lower than in the naive approach. This change only occurs on sections of the flux rope that deviate from a straight cylindrical geometry, which is apparent as both field lines do not differ at the start where the curvature is largely constant.

\begin{figure*}[ht!]
\center
\includegraphics[trim=10 20 5 10, clip,width=.98\linewidth]{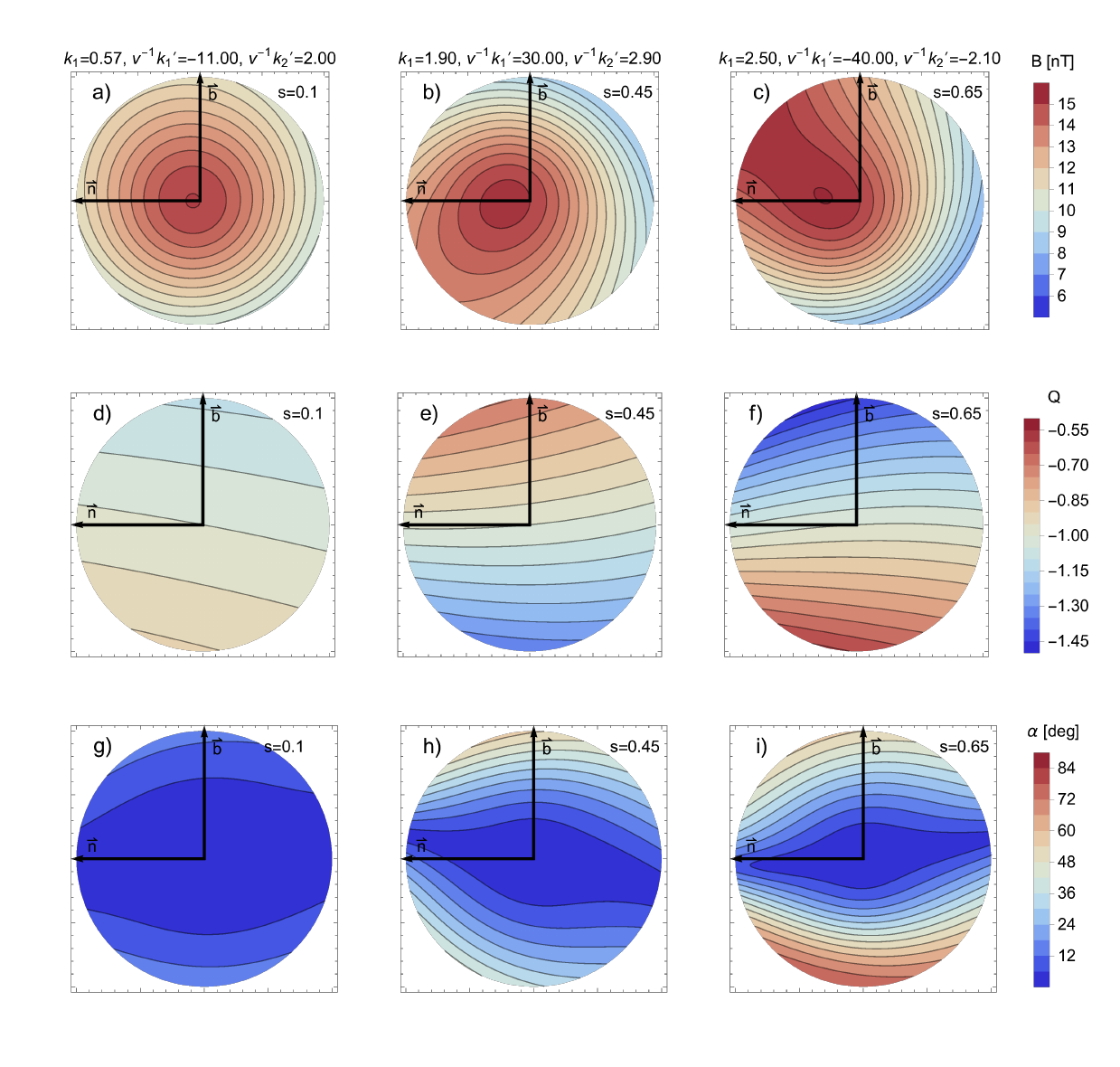}
\caption{\label{fig:cross_sections} Cross-section plots for the total magnetic field strength $B$ (a-c), the local twist $Q$ (d-f) and the misalignment angle $\alpha$ between the current density and the magnetic field vectors (g-i) at three different positions along our exemplary flux rope structure. The positions are also marked as orange points in Figure \ref{fig:exemplary_flux_rope}. The two black arrows correspond to the $\vb*{n}$ and $\vb*{b}$ Frenet-Serret vectors and are also marked as such. On the left side the flux rope is nearly toroidal with a very small curvature value. The middle and right plots represent sections of the flux rope with higher curvature and also a higher rate of change in curvature leading to apparent rotations of the cross-section profile.}
\end{figure*}

Figure \ref{fig:cross_sections}a-c shows the cross-sections for the total magnetic field strength $B$ at three different positions along the flux rope axis $(s=0.1\,s=0.45$ and $s=0.65)$. These three positions are also marked as orange dots in Figure \ref{fig:exemplary_flux_rope}. For these plots we use a specific parallel transport frame so that $\vb*{n}_1=\vb*{n}$ and $\vb*{n}_2=\vb*{b}$. The frame then locally coincides with the Frenet-Serret frame which simplifies the geometrical interpretation. For $s=0.1$ (a) we see that the profile very closely matches the cross-section of a classical uniformly twisted torus model with a small shift of the maximum towards the inner section of the curved flux rope \cite<e.g.,>[]{Vandas_2017}. In both cases for $s=0.45$ (b) and $s=0.65$ (c) we see that the curvature is more than twice as large so that the intensity profile is shifted further back. A new feature in these results (b-c) is now what appears to be a rotation of the entire profile. For $s=0.45$ (b) the profile appears to be rotated counter-clockwise while for $s=0.65$ (c) the rotation is in the clock-wise direction. The difference in between these two cases is the sign of $\partial_s k_1$ so that the curvature is increasing for $s=0.45$ where it is decreasing for $s=0.65$. This is a consequence of the $\varphi$ dependency of the extra terms in the general expression for $B^\varphi$. This rotation will also persist in the case the flux rope is constrained to a plane, meaning that an arbitrarily curved flux rope in a plane cannot be symmetric with respect to up and down as a perfect toroidal flux rope would be.

Figure \ref{fig:cross_sections}d-f shows cross-section for the local twist $Q$. They show that the local twist profile changes drastically when the curvature changes, specifically at positions near $\varphi=\nicefrac{\pi}{2}$ and $\varphi=\nicefrac{3\pi}{2}$. As in the intensity plots (b-c), the twist cross-sections for (e-f) are also inversed with respect to each other. In the most extreme case (f), the magnetic field twist changes by a factor of almost $30\%$ compared to the reference value for zero curvature. Despite the appearance of a balance for the size of the higher and lower twist regions the field line in our example exhibits a lower twist. This is due to the fact that a field line will azimuthally rotate faster when experiencing a higher twist, which leads it to occupy the lower twist region for longer. As a result, any field line in our model will always have a lower total twist number per unit length than would be expected from a toroidal or cylindrical model.

Figure \ref{fig:cross_sections}g-i shows cross-section for the misalignment angle $\alpha$ between the current density and magnetic field vector as proxy for the Lorentz force. We find that for all examples the flux rope is largely force-free within the core but no longer force-free near the flux rope boundary. The result for $s=0.1$ (g) is similar to the result from Figure 7 in \citeA{Vandas_2017}, except that the entire profile is rotated by almost 90$^\circ$. For both the cases $s=0.45$ (h) and $s=0.65$ (i) the Lorentz forces become very strong at the boundaries near $\varphi=\pm \nicefrac{\pi}{2}$. These plots can also be related to our approximate result that we computed for the net radial Lorentz force in Eq. \eqref{eq:lorentz_force_net_radial}, where we see that this net force primarily acts along the $\pm \vb*{n}_2$ vector \add{because $\partial_s k_1$ is much larger than both $k_1$ and $\partial_s k_2$}. \remove{and that the hoop force and tension force must be comparably small.}

\begin{figure*}[ht!]
\center
\includegraphics[trim=35 10 35 10, clip,width=.95\linewidth]{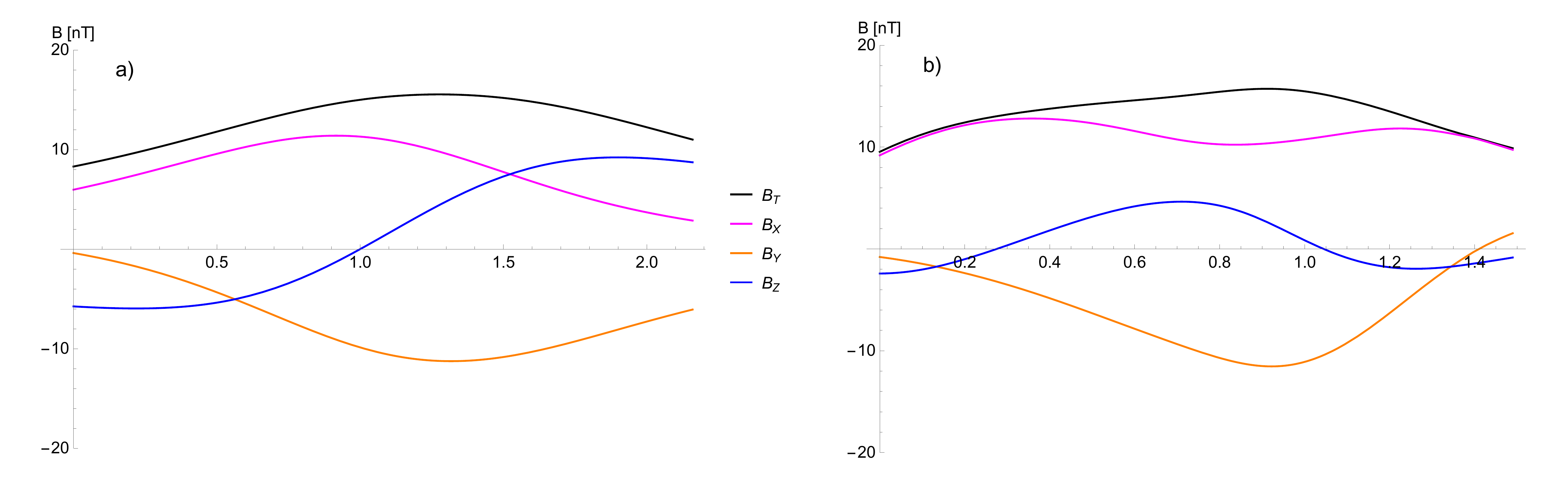}
\caption{\label{fig:model_profiles} Two synthetic in situ profiles generated by the virtual spacecraft trajectories $t_1$ and $t_2$ given in Figure \ref{fig:exemplary_flux_rope}. The profile generated by trajectory $t_1$ (a) represents a classical frontal encounter with a toroidal flux rope structure generates the commonly seen rotating magnetic field profile for a curved torus geometry. The profile generated by trajectory $t_2$ (b) describes a flanking pass where the spacecraft trajectory is largely parallel to the flux rope axis most of the time.}
\end{figure*}

Figure \ref{fig:exemplary_flux_rope} also shows two virtual spacecraft trajectories $t_1$ and $t_2$, as black lines with arrows giving the direction, from two different vantage points. The first virtual trajectory $t_1$  represents a more or less classical toroidal magnetic flux rope measurement. In the second case, for the trajectory $t_2$, the spacecraft traverses side ways through a large portion of the flux rope. The resulting synthetic in situ magnetic field profiles, for both cases, are shown in Figure \ref{fig:model_profiles} and are plotted in terms of an arbitrary length measure. For the first case we see that the profile matches the classical case of a rotating magnetic field profile with unipolar $B_{X/Y}$ components and a dipolar $B_Z$ component. The latter part of the synthetic measurement, which corresponds to the inner region of the curved flux rope, has a stronger absolute magnetic field strength creating an asymmetric profile. Our second case is very different and does not show the magnetic field rotation that you would expect from a typical flux rope. Such unusual flux rope signatures can also sometimes be seen in real in situ data and could \remove{therefore} be interpreted as \change{flank hits that are}{a flank encounter that is} similar to our proposed scenario \cite<e.g.,>[]{Marubashi_2007,moestl_2010, Owens_2012}.

\section{Discussion \& Conclusion} \label{sec:dc}

In this paper, we have introduced the mathematical concepts that are necessary to describe writhed flux rope structures under the constraint of a circular cross-section and conserved axial flux. We furthermore derived \change{specific}{a class of} solutions to the magnetic field equations for this geometry which requires the introduction of an implicitly defined radial current that imposes further conditions on our flux rope model that we do not investigate in detail. \add{While} the flux rope still possesses a clear magnetic boundary at its edge\change{but not for the current, as  $j^r_c \neq 0$}{, the same cannot be said for the current}. \change{and this implicitly sets unknown conditions on the external region that surrounds it.}{This sets unknown conditions on the external region that surrounds the flux rope.} As was already remarked in NC18, radial currents can also be introduced into cylindrical or toroidal geometries to remove the azimuthal symmetry. The exact nature of such a radial current and the resulting physical implications are, at this, not entirely clear to us. Additionally, it may be possible to find more consistent solutions by allowing for a varying cross-section that can be locally adapted to define a boundary through which there is no magnetic flux and no current. A very special case of such a flux rope, for a closed path $\vb*{\gamma}(s)$ that is confined to a plane and is strictly convex, is given \change{in}{by} \citeA{Yeh_1986}.

In the limit of constant curvature values, the presented flux rope model is reduced to the classical cylindrical or toroidal flux rope models. In such a simpler configuration it is thus possible to fully reproduce any reconstruction using a purely cylindrical or toroidal flux rope model regardless of a specific magnetic field configuration. In some cases, it may then be possible to further improve the reconstruction by slightly perturbing the flux rope axis in order to better match the magnetic field measurements. \add{The latter terms in Eq.} \eqref{eq:bp_result} \add{scale with an additional factor of $r$ compared to the primary term so that the magnetic field, in curvilinear coordinates, will mostly change near the flux rope boundaries. The above mentioned technique could thus be used to improve reconstructions near the start or end of observations as these regions are often not well described by contemporary models.}
At the current time, there are to us no known tools that would make it possible to directly infer the full global structure of a writhed flux rope just from the in situ magnetic field measurements. \add{The additional degrees of freedom that are introduced in our approach, by allowing a flexible flux rope axis $\vb*{\gamma}$, suggests that this is a monumental task and way beyond current standard fitting methods. Our previous studies} \cite<e.g.,>[]{Weiss_2021a, Weiss_2021b} \add{show that reconstructions are problematic for far simpler geometries even when using multiple spacecraft at smaller separations. Separate studies }  \cite<e.g.,>[]{Haddad_2011,Haddad_2019} \add{indicate that it might be hard to differentiate between a succession of writhed magnetic field lines and a twisted flux rope.}

\remove{The latter terms in Eq. scale with an additional factor of $r$ compared to the primary term so that the magnetic field, in curvilinear coordinates, will mostly change near the flux rope boundaries. It can thus be expected that the difference for in situ profiles between our approximate and naive solution is only marginal. The largest effect will be that our writhed model allows us to change the geometry of the flux rope axis.}

With the usage of Legendre polynomials and NC16/NC18 there are now at least two approximate approaches for describing the internal magnetic field structures for the type of models that we use in this paper. For uniformly twisted fields the Legendre approach is clearly superior but it may have unknown problems in other scenarios. We also have not attempted to describe the evolution of the coefficients regarding flux rope expansion which will differ depending on the polynomial basis that is used. In the future it may be necessary to more closely analyze these approaches for different scenarios. In both cases, the degrees of freedom that arise when using many coefficients are too high to be properly used for real \change{data}{scenarios}. It is thus clear that no matter which approach is used that there must be a simple description of the magnetic field. This can either be a uniform twist number as with the uniformly twisted field, an $\alpha$ parameter for a linear force-free field, or another parameter for \change{another type of}{a different} distribution. \add{The types of distributions that should be used are still under debate} \cite<e.g.,>[]{Wang_2018,Vandas_2019,Pal_2021}. \add{An interesting key result of our new model is that the twist per unit length will effectively decrease due to writhing of the flux rope. This has the consequence that using cylindrical or toroidal approximations will lead to overestimates of the twist when applied to in situ measurements of writhed flux rope structures.}

The model is fairly straightforward to implement for numerical applications. The parametrized path $\vb*{\gamma}(s)$ can easily be described using splines, which must be of at least third order. \add{Using fourth order splines, as we used for our exemplary flux rope, guarantees that $k_{1/2}$ are smooth.} Transformation of Cartesian coordinates into our curvilinear coordinate system is simple by using minimization algorithms under the condition that good initial starting values are used (see Section \ref{sec:impl}). The spline implementation also provides a way to change the overall geometry of the curve by moving the control nodes without changing key properties of the flux rope such as the magnetic flux. This may allow us, in the future, to build a highly simplified analytic simulation and evolve the underlying flux rope axis over time according to the arising net Lorentz forces or additional external factors such as the solar wind drag force or the ambient coronal magnetic field. 

If we want to consider time-dependent changes we also need to consider flux rope expansion or distortions of the cross-section. The results from Figure \ref{fig:cross_sections}g-i show highly asymmetric Lorentz force distributions when we include effects due to changing curvature or torsion. It is therefore highly likely that our usage of a circular cross-section is a strong approximation and that this shape will become additionally distorted over time. The arising Lorentz forces do not necessarily lead to an additional expansion and the flux rope expansion is still expected to be dominated by the pressure gradient at higher distances from the sun. To test how valid the circular cross-section approximation is, and under which conditions it may be suitable, would require more sophisticated numerical MHD simulations \cite<e.g.,>[]{ Scolini_2021,Lynch_2022}.


The presented flux rope model, making use of the spline approach, can also be readily implemented in a forward simulation model. An approach similar to the one used in \citeA{Hinterreiter_2021} could be used to drive the changes in the flux rope geometry due to interactions with the ambient solar wind. Solar wind velocity maps, for the inner heliosphere, can be generated by simulations such as Enlil \cite{Odstrcil_2003}, HUXt \cite{owens_2020} or THUX \cite{Reiss_2020}. Assuming that the resulting forward simulations are fast enough, one could attempt to build a fitting pipeline using a Monte-Carlo approach as was done for simpler analytical flux rope models \cite<e.g.,>[]{Weiss_2021a}. \add{This approach could make up for the lack of a direct fitting method due to the complexity of the model. A big problem regarding such simulations would be the initial conditions for the geometry of the writhed flux rope. Accurate initial conditions would significantly accelerate any Monte Carlo based approach by limiting the degrees of freedom. Methods developed for other semi-empirical or empirical flux rope models }\cite<e.g.,>[]{Kay_2013,Palmerio_2017} \add{could potentially be adopted to estimate, or limit, the geometry of our flux ropes within our paradigm when close to the Sun.}


Future studies are planned in which we aim to further develop this flux rope model by extending the cross-section geometry to either an elliptic shape or generally distorted shapes. Implementing a distortion factor $\delta$ into Eq. \eqref{eq:coordinate_system}, as was done in NC18, would achieve this but we would not have control over the orientation of the cross-section as the normal vectors evolve with the parametrized curve. In NC18 this orientation problem does not exist as the major axis of the ellipse is aligned with respect to the $z$ axis. We would thus need to add additional degrees of freedom to control the non-axisymmetric cross-section, which will significantly increase the complexity of the model.

\section{Data Availability}

\add{The Mathematica notebook that was used to generate the figures in this manuscript is openly available:} \url{https://doi.org/10.5281/zenodo.7294481}

\acknowledgments
A.J.W, C.M., R.L.B, M.A.R. and T.A. thank the Austrian Science Fund (FWF): P31521-N27, P31659-N27. T.N-Ch acknowledges the NASA-GSFC Heliophysics Internal Fund (HIF) “Physics-driven modeling of the Interplanetary coronal mass ejections distortions”. A. J. W. and C. M. were funded by the European Union. Views and opinions expressed are however those of the author(s) only and do not necessarily reflect those of the European Union or the European Research Council Executive Agency. Neither the European Union nor the granting authority can be held responsible for them.

\bibliography{refs}

\end{document}